\documentclass[12pt]{article}
\usepackage{a4wide,epsfig,amsmath,amssymb,cite,scalefnt}
\usepackage{feynarts}

\newcommand{\lnMstu}{L_{\tilde{t}_1}}
\newcommand{\lnMstd}{L_{\tilde{t}_2}}

\newcommand{\lnMt}{L_t}

\newcommand{\Mgl}{m_{\tilde g}}

\newcommand{\Msbu}{m_{\tilde{b}_1}}
\newcommand{\Msbd}{m_{\tilde{b}_2}}
\newcommand{\Mstu}{m_{\tilde{t}_1}}
\newcommand{\Mstd}{m_{\tilde{t}_2}}
\newcommand{\Sq}{\sin 2 \theta_b}
\newcommand{\St}{\sin 2 \theta_t}
\newcommand{\Sqq}{\sin^2 2 \theta_b}
\newcommand{\Stq}{\sin^2 2 \theta_t}

\newcommand{\abbrev}{\scalefont{.9}}
\newcommand{\drbar}{$\overline{\mbox{\abbrev DR}}$}
\newcommand{\msbar}{$\overline{\mbox{\abbrev MS}}$}

\newcommand{\msbarmath}{\overline{\rm\abbrev MS}}

\begin{document}



\title{\vskip-3cm{\baselineskip14pt
    \begin{flushleft}
      \normalsize SFB/CPP-09-120 \\
      \normalsize TTP/09-45  
  \end{flushleft}}
  \vskip1.5cm
   ${\cal O}(\alpha_s^2)$ corrections to fermionic Higgs decays \\
  in the MSSM }
\author{ 
  L. Mihaila, C. Rei{\ss}er\\
  {\small\it Institut f{\"u}r Theoretische Teilchenphysik,}\\
{\small\it Karlsruhe Institute of Technology (KIT)}\\
  {\small\it 76128 Karlsruhe, Germany}\\
}

\date{}

\maketitle

\thispagestyle{empty}

\begin{abstract}
We compute the two-loop corrections of ${\cal O}(\alpha_s^2)$ to the 
 Yukawa couplings in the framework of the  Minimal
Supersymmetric Standard Model (MSSM). The calculation is performed using the 
effective Lagrangian approach under the approximation of neglecting the Higgs 
boson mass with respect to the top quark, gluino  and all squark flavour 
masses. As an application we derive the  ${\cal O}(\alpha_s^2)$ corrections
 to the partial decay width of the lightest Higgs boson to a bottom quark pair.
 We find that the two-loop 
corrections are sizable for large values of $\tan\beta$ and low CP-odd Higgs
 boson mass. With our calculation of  the ${\cal O}(\alpha_s^2)$ corrections
the remaining  theoretical uncertainties reduce  below a few percent.
\medskip

\noindent
PACS numbers: 11.30.Pb, 12.38.-t, 12.38.Bx, 12.10.Kt

\end{abstract}


\section{Introduction}
One of the main purposes of the CERN Large Hadron Collider (LHC) is the
search for  Higgs bosons. The discovery of a light Higgs boson is a
decisive test for all models predicting    supersymmetric (SUSY)
particles at the TeV scale and in particular of the 
MSSM~\cite{Nilles:1983ge}. A remarkable feature of the MSSM
is the restricted Higgs sector. This allows Higgs searches without any
assumption about the mechanism of SUSY breaking, but only constraints
from the Higgs sector~\cite{Carena:2002qg}. For Higgs boson searches
 at the hadron colliders, 
two new complementary
 benchmark scenarios, the ``small $\alpha_{\rm eff}$'' and the
 ``gluophobic'' scenarios,
have been proposed  in addition to  those used at the CERN Large
Electron-Positron Collider (LEP) for the MSSM
Higgs searches at hadron colliders.\\
More precisely, in the ``gluophobic'' scenario the gluon fusion
 process is strongly
suppressed due to cancellation between top quark and squark loop
contributions. Nevertheless, the channel $t\bar{t}\to t\bar{t} h\to
t\bar{t} b\bar{b}$ is enhanced as compared to the Standard Model (SM) case, 
so that this becomes the most promising  
detection mode. In the ``small $\alpha_{\rm eff}$'' scenario
 the decay width for $h\to b\bar{b}$ is much smaller than its SM
value. In this case the complementary channel $h\to\gamma\gamma$ is
enhanced as compared to the SM and it becomes   the preferred
detection mode.\\
 For a light Higgs boson ($m_h\le 130$~GeV)
the decay $h\to b\bar{b}$ is the dominant  mode, but its detection at
 hadron colliders 
is difficult due to large QCD backgrounds for the $b$
jets. However, at lepton colliders the Higgs boson search relies on $b$
 tagging that can be performed with high efficiency.

 For the discovery of the Higgs bosons, the cross section of the
 main production
channels, decay widths and branching ratios are necessary to be known
with high accuracy. 
 Within the SM the radiative corrections to the fermionic
 Higgs decay were intensively studied in the literature. The QCD and EW
 corrections are known up to the three-loop order: ${\cal O}(\alpha_s^3)$ 
originating from the light degrees of freedom  were first
 derived in Ref.~\cite{Chetyrkin:1996sr} and the top-induced 
 ${\cal O}(\alpha_s^3)$ corrections in Ref.~\cite{Chetyrkin:1997vj}; 
the QCD-EW interference contributions  of
${\cal O}(\alpha_s^2 x_t)$\footnote{For the definition of
 the parameter $x_t$ see Section~\ref{sec::hdecay}.} can be found
 in Ref.~\cite{Chetyrkin:1996ke}.
In this paper we concentrate on the fermionic 
Higgs decay in the MSSM, for moderate Higgs boson masses $M_h\le 130$~GeV. 
The processes $h\to b\bar{b}$ and $h \to \tau^+\tau-$ (the second most
important decay mode) are affected by large radiative corrections
 for scenarios with large values of $\tan\beta$ and moderate values of 
the neutral CP-odd Higgs boson mass $M_A$
($\tan\beta\ge 20\,, M_A\le 250$~GeV)~\cite{Hall:1993gn}. Apart
 from pure QCD and EW corrections mentioned above,
there are Higgs boson propagator corrections and vertex corrections due
to SUSY particles. The first class of radiative corrections, can be
taken into account by introducing the effective mixing angle
$\alpha_{\rm eff}$ that diagonalizes the neutral Higgs boson mass
 matrix~\cite{Heinemeyer:2000fa}.
 Such type of corrections are known analytically up to
 two-loop order in supersymmetric QCD (SQCD) and the supersymmetric
 electroweak theory (SEW), see \cite{FeynHiggs} and references
 therein. 
The   second class of radiative corrections are especially important for 
 large values of $\tan\beta$. Usually, they
are derived  using the 
 effective Lagrangian approach~\cite{Carena:1999py}. The one-loop
 contributions are known since long~\cite{Hall:1993gn,Guasch:2003cv},
 while the two-loop 
 corrections  have been computed very recently\cite{Noth:2010jy}.\\
It is the aim of this paper to present the complete two-loop SQCD corrections
to the decay width $h\to b\bar{b}$, taking into account the exact 
dependence on the supersymmetric particle masses and working in the
full theory.

The paper is organized as follows: in the next section we introduce our 
framework and the quantities required for the computation of the decay
 width $\Gamma(h\to q\bar{q})$ through two loops in the MSSM. In Section 
3 we describe the actual two-loop calculation, pointing out the connection
 between the radiative  corrections to the vertex $hq\bar{q}$ and those 
to the quark propagator through the low-energy theorem. The formalism discussed
in this section is valid for a general quark flavour. However, we 
specify our calculation for the case of bottom quark  which  generates the
 dominant decay  mode of a light Higgs boson.
 In Section 4 we
 perform a  numerical analysis and discuss the phenomenological implications   
of the two-loop SQCD vertex corrections to $\Gamma(h\to b\bar{b})$. We present the conclusions in
 Section 5.

\section{\label{sec::framework} Notation and theoretical framework}
 The  part of the MSSM  Lagrangian
 describing the fermionic Higgs decay  
can be written in the following form
\begin{eqnarray}
{\cal L}&=&{\cal L}_{\rm QCD}+ {\cal L}_{\rm SQCD} 
+\sum_{i=1,2}{\cal L}_{{\rm q}\phi_i} 
 + \sum_{i=1,2}{\cal L}_{{\rm \tilde{q}}\phi_i}\,
\end{eqnarray}
where 
\begin{eqnarray}
{\cal L}_{{\rm q}\phi_i}=-\sum_{q=1}^6 \frac{m_q}{v} g_q^{\phi_i}
 \bar{q} q \phi_i \quad \mbox{and} \quad
{\cal L}_{{\rm \tilde{q}}\phi_i}=-\sum_{q=1}^6\sum_{r,k=1,2} 
\frac{m_q}{v} g_{\tilde{q};kr}^{\phi_i}\tilde{q}^\star_{k}\tilde{q}_r \phi_i\,.
\end{eqnarray}
${\cal L}_{\rm QCD}+ {\cal L}_{\rm SQCD}$ denotes the supersymmetric
 extension of the full QCD Lagrangian with six quark flavours.
 The couplings $g_q^{\phi_i}$ and
 $g_{\tilde{q};kr}^{\phi_i}$ are defined in Table~\ref{tab::yukawa_coeff},
  $m_q$ denotes the mass of quark $q$,  $v=\sqrt{v_1^2+v_2^2}$ with
 $v_i\,,i=1,2$, the vacuum expectation values of the two Higgs
 doublets of the MSSM. The fields
$\tilde{q}_i\,,i=1,2$, denote the squark mass eigenstates, while
 $\theta_q$ stands for the mixing angle defined through:
\begin{eqnarray}
\sin 2\theta_q=\frac{2 m_q X_q}{m_{\tilde{q}_1}^2-m_{\tilde{q}_2}^2}\,,\quad 
X_q = A_q-\mu_{\rm SUSY} \bigg\{\begin{array}{ll}
\tan\beta\,, &\mbox{ for  down-type quarks}\\
\cot\beta\,,  &\mbox{ for  up-type quarks}
\end{array}  \,,
\label{eq::mixangle}
\end{eqnarray}
where $A_q$ is  the trilinear coupling and $\mu_{\rm SUSY} $    the
Higgs-Higgsino bilinear coupling.
The fields $\phi_i\,, i=1,2$, denote the neutral CP-even components
 of the MSSM Higgs
 doublets and they are related  to the Higgs mass eigenstates
 through the orthogonal
 transformation
\begin{eqnarray}
\left(
\begin{array}{c}
H\\h
\end{array}
\right) =
\left(
\begin{array}{cc}
\cos\alpha&\sin\alpha\\
-\sin\alpha & \cos\alpha 
\end{array}
\right)
\left(
\begin{array}{c}
\phi_1\\
\phi_2
\end{array}
\right)\,.
\label{eq::hmix}
\end{eqnarray}
As usual, $h$ stands for the lightest Higgs boson. 
The mixing angle $\alpha$ is determined at the leading order through
\begin{eqnarray}
\tan 2\alpha= \tan 2 \beta \frac{M_A^2+M_Z^2}{M_A^2-M_Z^2}\,; 
\quad -\frac{\pi}{2}<\alpha<0\, ,
\label{eq::h_alpha}
\end{eqnarray}
where $M_Z$ is the mass of the Z boson and $\tan  \beta=v_2/v_1$.

\begin{table}
  \begin{center}
    \begin{tabular}{|l||c|c|c|c|}
\hline
f&$g_q^{\phi_1}$&$g_{\tilde{q};11}^{\phi_1}$&$g_{\tilde{q};12}^{\phi_1}=
g_{\tilde{q};21}^{\phi_1}$&$g_{\tilde{q};22}^{\phi_1}$\\
\hline
up&$0$&$-\mu S_q/S_\beta$&$-\mu  C_q/S_\beta$ &$\mu S_q/S_\beta$\\
\hline
down&$1/C_\beta$& $(2 m_q+A_q S_q)/C_\beta$&$
A_q C_q/C_\beta$&$(2 m_q-A_q S_q)/C_\beta$\\
\hline
\hline
f&$g_q^{\phi_2}$&$g_{\tilde{q};11}^{\phi_2}$&$
g_{\tilde{q};12}^{\phi_2}=g_{\tilde{q};21}^{\phi_2}$&$
g_{\tilde{q};22}^{\phi_2}$\\
\hline
up&$1/S_\beta$& $(2 m_q+A_q S_q)/S_\beta$&$A_q C_q/S_\beta$&$
(2 m_q-A_q S_q)/S_\beta$\\
\hline
down&$0$&$-\mu S_q/C_\beta$&$-\mu  C_q/C_\beta$ &$\mu S_q/C_\beta$\\
\hline
    \end{tabular}
    \caption{\label{tab::yukawa_coeff}Yukawa coupling coefficients for up
      and down type quark and squark, where
$S_q=\sin 2\theta_q$ and $C_q=\cos 2 \theta_q$, and $S_\beta=\sin \beta$ 
and $C_\beta=\cos  \beta$.}
  \end{center}
\end{table}

In the following, we assume the mass of the lightest Higgs boson
$h$ to be much
 smaller than the mass of the top-quark and of the SUSY particles, as well as
  all the other Higgs bosons. In this case, the physical phenomena
 at low energies can be described with an 
effective theory containing  five  quark flavours and the light Higgs.
  At leading order in the heavy masses, the  effective Lagrangian
 ${\cal L}_Y^{\rm eff}$
can be written as a linear combination of three physical
 operators~\cite{Spiridonov:1984,Chetyrkin:1997un} constructed from the
 light degrees of freedom
\begin{eqnarray}
{\cal L}\longrightarrow {\cal L}_Y^{\rm eff} + {\cal L}_{\rm QCD}^{(5)}\,;\quad
{\cal L}_Y^{\rm eff}  = -\frac{h^{(0)}}{v^{(0)}}\left[
C_1^0{\cal O}_1^0 + \sum_{q}\left( C_{2q}^0{\cal O}_{2q}^0
 + C_{3q}^0{\cal O}_{3q}^0\right)
\right]\,,
\label{eq::eft}
\end{eqnarray}
where $ {\cal L}_{\rm QCD}^{(5)}$ denotes the Lagrangian of QCD with five active
 flavours and the coefficient functions $C_i\,, i=1,2q,3q$, parametrize the
 effects of the heavy particles on the low-energy phenomena.
 The superscript
 $0$ labels bare quantities. The three
 operators are defined as

\begin{eqnarray}
{\cal O}_1^0 &=& (G_{\mu,\nu}^{0,\prime,a})^2\,,\nonumber\\
{\cal O}_{2q}^0 &=& m_q^{0,\prime}\bar{q}^{0,\prime}
q^{0,\prime}\,,\nonumber\\
{\cal O}_{3q}^0 &=& \bar{q}^{0,\prime}(i\,/\!\!\!\! D^{0,\prime}
-m_q^{0,\prime})q^{0,\prime}\,,
\label{eq::ops}
\end{eqnarray}
where $G_{\mu,\nu}^{0,\prime,a}$ and $ D_{\mu}^{0,\prime}$ are the
gluon field strength tensor and the covariant derivative, respectively, and 
the primes label the quantities in the effective theory. 
 The operator ${\cal O}_{3q}$
 vanishes by the fermionic equation
 of motion and it will not contribute to physical observables.
 So, the last term in Eq.~(\ref{eq::eft})  might be omitted,
 once the coefficients $C_1^0, C_{2q}^0$ are determined.
The coefficient functions contain  information about the heavy particles
 that were integrated out in the construction of the effective theory.
On the contrary, as can be understood from Eq.~(\ref{eq::ops}), the
operators encounter only the effects of the light degrees of freedom. 

The relations between the
parameters and fields in the full and effective theories are
given by
\begin{eqnarray}
G_{\mu,\nu}^{0,\prime,a}&=&(\zeta_3^{(0)})^{1/2}
G_{\mu,\nu}^{0,a}\,,\nonumber\\
q^{0,\prime}&=& (\zeta_2^{(0)})^{1/2} q^{0}\,,\nonumber\\
g_s^{0,\prime}&=& \zeta_g^{(0)} g_s^{0}\,,\nonumber\\
m_q^{0,\prime}&=&\zeta_m^{(0)}m_q^{0}\,,
\label{eq::dec}
\end{eqnarray}
where $g_s=\sqrt{4\pi\alpha_s}$ is the strong coupling. The coefficients
$\zeta_3^{(0)}\,,\zeta_2^{(0)}\,,\zeta_g^{(0)}\,,\zeta_m^{(0)}$ are the bare
decoupling coefficients. They may be computed from the transverse part
 of the gluon polarization  function and the vector and scalar
part of the quark self-energy  via~\cite{Chetyrkin:1997un}
\begin{eqnarray}
\zeta_3^{(0)}=1+\Pi^{0,h}(0)\,,\nonumber\\
\zeta_2^{(0)}=1+\Sigma_v^{0,h}(0)\,,\nonumber\\
\zeta_m^{(0)}=\frac{1-\Sigma_s^{0,h}(0)}{1+\Sigma_v^{0,h}(0)}\,.
\label{eq::letse}
\end{eqnarray}
For the derivation of the coefficient $\zeta_g^{(0)}$ one has to
consider in addition one vertex involving the strong coupling, for
example $\bar{q}qg$ or $\bar{c}cg$, where $c$ denotes the Faddeev-Popov ghost.
The decoupling coefficients are independent of the momentum transfer, so
that they can be evaluated at vanishing external momenta. The
superscript $h$ indicates that in the framework of Dimensional
Regularization (DREG) or Dimensional Reduction (DRED) only diagrams
containing at least one heavy particle inside the loops contribute and that
 only the hard regions in the asymptotic expansion of the diagrams are
 taken into account. They have been computed in QCD including
 corrections up to the four-loop
order  for the strong coupling~\cite{Schroder:2005hy} and 
three-loop order for quark masses~\cite{Chetyrkin:1997un}.
In the MSSM the  two-loop 
 SQCD~\cite{Harlander:2005wm, Bednyakov:2007vm,Bauer:2008bj}
 and SEW~\cite{Bednyakov:2009wt} expressions are  known.
Similar to the case of SM, the decoupling coefficients derived within
the MSSM can be  connected through the Low Energy Theorem
 (LET)~\cite{Ellis:1975ap} with
the coefficients $C_1^0, C_{2q}^0$. We discuss in more detail the relation
between the  coefficients $C_{2q}^0, C_{3q}^0$ 
 and $\zeta_m^{(0)}, \zeta_2^{(0)}$ in Subsection~\ref{sec::letsusy}.\\

The renormalization procedure of the dimension four operators 
in the Minimal Subtraction Scheme  within DREG  (\msbar{})~\cite{Bardeen:1978yd}
 is known since long
 time~\cite{Spiridonov:1984}. The main aspect is that different operators in
 general mix under renormalization. For the convenience of the reader we
 reproduce  the results for the renormalization constants of the
 operators ${\cal O}_1^0$ and ${\cal O}_{2q}^0$ that are of interest for
 the fermionic Higgs decays
\begin{eqnarray}
&&{\cal O}_1= Z_{11} {\cal O}_1^0 +Z_{12} {\cal O}_{2q}^0\,,\qquad
 {\cal O}_2= Z_{22}{\cal O}_{2q}^0\,,\quad \mbox{where} \nonumber\\
&&Z_{11}=
\left(1-\frac{\pi}{\alpha_s^\prime}\frac{\beta(\alpha_s^\prime)}
{\epsilon}\right)^{-1}
, \,\,\, Z_{12}=-\frac{4\gamma_m(\alpha_s^\prime)}{\epsilon}
\left(1-\frac{\pi}{\alpha_s^\prime}\frac{\beta(\alpha_s^\prime)}
{\epsilon}\right)^{-1},
\, Z_{22}=1\,,\\
&&C_1= Z_{11}^{-1} C_1^0 \,,\qquad\qquad\qquad C_{2q}=C_{2q}^0-
\frac{Z_{12}}{Z_{11}} C_1^0 \,.
\end{eqnarray}
The explicit expressions for the $\beta$-function and quark mass anomalous
dimension $\gamma_m$ of QCD with $n_l=5$ active flavours at the one-loop
order that are needed in the present paper, are given by
\begin{eqnarray}
\beta(\alpha_s^\prime)&=&-\left(\frac{\alpha_s^\prime}{\pi}\right)^2\beta_0
 +{\cal  O}((\alpha_s^\prime)^3)\,,\quad \beta_0 =\frac{11}{4}
-\frac{n_l}{6}\,,\nonumber\\
\gamma_m(\alpha_s^\prime) &=& -\frac{\alpha_s^\prime}{\pi} \gamma_0 +{\cal
 O}((\alpha_s^\prime)^2)\,,\qquad\quad \gamma_0= \frac{3}{4} C_F\,.
\label{eq::andim}
\end{eqnarray}
The bare coefficient functions $C_i^0\,, i=1,2q$,  must be computed
diagrammatically. For the calculation of the  ${\cal O}(\alpha_s^2)$
 corrections to the process
$h\to q \bar{q}$,  the knowledge of the coefficient functions
$C_1^0$  and  $C_{2q}^0$  is required at the one- and two-loop order,
 respectively. 

The renormalized coefficient functions and operators are finite but not
renormalization group (RG) invariant. In Ref.~\cite{Chetyrkin:1996ke},
 a redefinition
of the coefficient functions and operators was introduced so that they
are separately RG invariant. This procedure allows us to choose independent
renormalization scales for coefficient functions and operators. In
practice, one makes a separation of  scales: one chooses $\mu\approx M_h$ 
for the
renormalization scale of the operators and  $\mu\approx \tilde{M}$
(where $\tilde{M}$ denotes an averaged mass for the heavy 
supersymmetric particles) for the coefficient functions. 
The new coefficient functions read~\cite{Chetyrkin:1996ke} 
\begin{eqnarray}
{\cal C}_1(\tilde{M}, M_h)&=&
\frac{\alpha_s^\prime(\tilde{M})\beta^{(5)}(\alpha_s^\prime(M_h))}
{\alpha_s^\prime(M_h)\beta^{(5)}(\alpha_s^\prime(\tilde{M}))}
 C_1(\tilde{M})\,,\nonumber\\
{\cal C}_2(\tilde{M}, M_h)&=& \frac{4 \alpha_s^\prime(\tilde{M})}{\pi
  \beta^{(5)}(\alpha_s^\prime(\tilde M))}[\gamma_m^{(5)}(\alpha_s^\prime(\tilde{M}))
-\gamma_m^{(5)}(\alpha_s^\prime(M_h))]C_1(\tilde{M})+   
C_{2q}(\tilde{M})\,,
\end{eqnarray}
where the superscript $(5)$ marks that $n_l=5$ in Eq.~(\ref{eq::andim}).
We employ this approach for the evaluation of the decay width 
   $\Gamma(h\to \bar{q}q)$. More details about the practical calculation
   are discussed in the next section.

\subsection{\label{sec::hdecay} Higgs decay width}
Once the renormalized coefficient functions ${\cal C}_1, {\cal C}_2$ are
known, the decay width for the process $h\to \bar{q}q$ can be predicted.
From Eqs.~(\ref{eq::eft}) and (\ref{eq::ops}) one can derive a general
formula for the inclusive $h\to \bar{q}q$ decay width~\cite{Chetyrkin:1996ke} 
\begin{eqnarray}
\Gamma(h\to \bar{q}q) = \Gamma^{(0)}(1+\bar{\delta}_u)^2\bigg[
(1+\Delta_q^{\rm QCD}) {\cal C}_2^2+ \Xi_q^{\rm QCD}{\cal C}_1{\cal C}_2
\bigg]\,,
\label{eq::gamma_gen}
\end{eqnarray}
where $\Gamma^{(0)}$ represents the complete leading order (LO) result given by
\begin{eqnarray}
\Gamma^{(0)} &=& \frac{N_c G_F M_h m_q^2}{4\pi\sqrt{2}}
\left(1-\frac{4m_q^2}{M_h^2}\right)^{3/2}\,.
\label{eq::born}
\end{eqnarray}
As is well known, the large logarithms of the type
 $\ln(M_h^2/m_q^2)$
can be resummed by taking $m_q$ in Eq.~(\ref{eq::born}) to be the
\msbar{} mass $m_q^{\msbarmath}(\mu)$ evaluated at the scale
$\mu=M_h$. The QCD correction $\Delta_q^{\rm QCD}$ is known since long 
time~\cite{Gorishnii:1990zu},
\begin{eqnarray}
\Delta_q^{\rm QCD} &=&
\frac{\alpha_s^{\prime}(\mu)}{\pi}\left(\frac{17}{3} + 2
  \ln\frac{\mu^2}{M_h^2}\right)\nonumber\\
&+&
\left(\frac{\alpha_s^{\prime}(\mu)}{\pi}\right)^2\bigg[
\frac{8851}{144}-\frac{47}{6}\zeta(2)-\frac{97}{6}
\zeta(3)+\frac{263}{9}\ln\frac{\mu^2}{M_h^2}
+ \frac{47}{12}\ln^2\frac{\mu^2}{M_h^2}
\bigg]\,,
\end{eqnarray}
with $\zeta(x)$ being the Riemann's zeta function.
The additional QCD correction   generated  through double-triangle
topologies $\Xi_q^{\rm QCD}$ was first computed in
Ref.~\cite{Chetyrkin:1996ke},
\begin{eqnarray}
 \Xi_q^{\rm QCD} &=& \frac{\alpha_s^{\prime}(\mu)}{\pi} {C_F} \left(-19
 +6\zeta(2)
-\ln^2\frac{m_q^2}{M_h^2} -6 \ln\frac{\mu^2}{M_h^2}\right)\,.
\end{eqnarray}
The universal corrections $\bar{\delta}_u$ of ${\cal
  O}(\alpha_s^n x_t)$,
 where $x_t=(\alpha_t/4\pi)^2=G_F M_t^2/(8\pi^2\sqrt{2})$, with
 $\alpha_t$ the top-Yukawa coupling, contain
the contributions from the renormalization of the Higgs wave function
and the vacuum expectation value~\cite{Kwiatkowski:1994cu},
 \begin{eqnarray}
\bar{\delta}_u &=& x_t\left[\frac{7}{2} +
  \frac{\alpha_s^{\prime}(\mu)}{\pi}\left(\frac{19}{3} -2\zeta(2)
+7\ln\frac{\mu^2}{M_t^2} \right) +{\cal O}(\alpha_s^2)
\right]\,.
\end{eqnarray}
The  coefficient functions $C_1\,, C_{2q}$ (and
implicitly ${\cal C}_1, {\cal C}_2$) 
 are known within SQCD at the one-loop order since quite some
 time~\cite{Spira:1993bb,Guasch:2003cv}. For completeness, we display
 them here 
 providing also ${\cal O}(\epsilon)$ terms that are necessary for the
 two-loop calculation.
\begin{eqnarray}
C_1&=&-\frac{\alpha_s^\prime(\mu)}{12\pi}\Bigg\{-\frac{\sin\alpha}{\cos\beta}
\Bigg[\frac{M_t^2\mu_{\rm SUSY} X_t}{4\Mstu^2\Mstd^2\tan\beta}
-\epsilon \frac{M_t\mu_{\rm SUSY}\sin
  2\theta_t}{8\tan\beta}\left(\frac{\lnMstu}{\Mstu^2} -
  \frac{\lnMstd}{\Mstd^2}\right) \Bigg]
\nonumber\\
&&+\frac{\cos\alpha}{\sin\beta}\Bigg[
\frac{4\Mstu^2\Mstd^2 + \Mstu^2 M_t^2 + \Mstd^2 M_t^2 - A_t M_t^2
  X_t}{4\Mstu^2\Mstd^2}\nonumber\\
&&
+\epsilon \frac{A_t M_t 
\sin 2\theta_t}{8}\left(\frac{\lnMstu}{\Mstu^2} -
    \frac{\lnMstd}{\Mstd^2}\right) +
\epsilon\frac{M_t^2}{4}\left(\frac{4 \lnMt}{M_t^2} 
+ \frac{\lnMstu}{\Mstu^2} + \frac{\lnMstd}{\Mstd^2}\right)
\Bigg]
\Bigg\}\,,
\\
C_{2q}&=&-\frac{\sin\alpha}{\cos\beta}\frac{1+\frac{\alpha_s^\prime(\mu)}{2\pi} C_F
A_b \Mgl\bigg[F_1(\Msbu^2,\Msbd^2,\Mgl^2)+\epsilon
F_2(\Msbu^2,\Msbd^2,\Mgl^2)\bigg]}{1+\frac{\alpha_s^\prime(\mu)}{2\pi} C_F
X_b \Mgl\bigg[F_1(\Msbu^2,\Msbd^2,\Mgl^2)+\epsilon
F_2(\Msbu^2,\Msbd^2,\Mgl^2)\bigg]}
\nonumber\\
&&
+\frac{\cos\alpha}{\sin\beta}
\frac{\frac{\alpha_s^\prime(\mu)}{2\pi} C_F
(-\mu_{\rm SUSY}\tan\beta) \Mgl\bigg[F_1(\Msbu^2,\Msbd^2,\Mgl^2)+\epsilon
F_2(\Msbu^2,\Msbd^2,\Mgl^2)\bigg]}{1+\frac{\alpha_s^\prime(\mu)}{2\pi} C_F
X_b \Mgl\bigg[F_1(\Msbu^2,\Msbd^2,\Mgl^2)+\epsilon
F_2(\Msbu^2,\Msbd^2,\Mgl^2)\bigg]}
\,,
\end{eqnarray}
where $C_F=4/3$, $L_x=\ln (\mu^2/m_x^2)$  
and the functions $F_1$ and $F_2$ are defined through
\begin{eqnarray}
F_1(x,y,z)&=&-\frac{x y
  \ln\frac{y}{x}+yz\ln\frac{z}{y}+zx\ln\frac{x}{z}}{(x-y)(y-z)(z-x)}\,,
\nonumber\\
F_2(x,y,z)&=&-\frac{1}{(x-y)(y-z)(z-x)}
\Bigg[x y
  \ln\frac{y}{x}(1+\ln\frac{\mu^2}{\sqrt{xy}})
\nonumber\\
&&
+yz\ln\frac{z}{y}(1+\ln\frac{\mu^2}{\sqrt{yz}})
+zx\ln\frac{x}{z}(1+\ln\frac{\mu^2}{\sqrt{xz}})\Bigg]\,.
\end{eqnarray}
In the above formulas, $\alpha_s^\prime(\mu)$ denotes the strong
 coupling constant computed in the \msbar{} scheme and taking into account 
$n_l=5$ active quark flavours. \\ 
The computation of the coefficient function $C_{2q}$ through two loops in SQCD
is discussed in some detail in the next section.

\section{\label{sec::nnlo}Calculation of the coefficient function
  $C_{2q}$ at next-to-next-to-leading order (NNLO) } 
 For the derivation of the coefficient functions one has to compute
Green functions in the full and effective theory and make use of 
the decoupling relations Eq.~(\ref{eq::dec}) to connect
them~\cite{Chetyrkin:1997un}.  
In general, one
Green function contains several coefficient functions. For example, 
 the amputated Green function
involving the $q \bar{q}$ pair and the zero-momentum insertion of the
operator ${\cal O}_h$ which mediates the couplings to
the light Higgs boson $h$ contains both coefficient functions $C_{2q}$
and $C_{3q}$.   
Similarly,  one possibility to compute  the coefficient function $C_1$ 
involves the 
Green function formed by the coupling of the operators ${\cal O}_h$
to two gluons.

 In the following, we restrict the discussion to the 
computation of the
 coefficient function $C_{2q}$. Considering the appropriate 
 one-particle-irreducible (1PI) Green function, we get
\begin{eqnarray}
\Gamma_{\bar{q}q{\cal O}_h}^{0}(p,-p)=i^2\int {\rm d}x {\rm d}y e^{i
  p (x-y)}\langle T q^{0}(x) \bar{q}^0(y){\cal
  O}_h(0)\rangle^{\rm 1PI}\,,
\end{eqnarray}
where $p$ is the outgoing momentum of $q$. In a next step we express
the operator ${\cal O}_h$ with the help of  Eqs.~(\ref{eq::eft}) and
(\ref{eq::ops}) and make use of the decoupling relations
Eq.~(\ref{eq::dec}).  One can  easily see that  the above Green function
 will get contributions only from the operators ${\cal O}_{2q}$ and
 ${\cal O}_{3q}$
 \begin{eqnarray}
\Gamma_{\bar{q}q{\cal O}_h}^{0,h}(p,-p)&=&-\zeta_2^{(0)}\int {\rm d}x {\rm d}y
 e^{i p (x-y)}\langle T q^{\prime,0}(x)
\bar{q}^{\prime,0}(y)( C_{2q}{\cal O}_{2q} +C_{3q}{\cal
  O}_{3q})\rangle^{\rm 1PI}\,,\nonumber\\
&=&\zeta_2^{(0)}\zeta_m^{(0)}(C_{2q}^{0}-C_{3q}^{0})m_b^{0}
+\zeta_2^{(0)}C_{3q}^{0}/\!\!\!p\,.
\label{eq::gf}
\end{eqnarray}
In the last step we have used the Feynman rules for the scalar dimension
four operators that can be found in Ref.~\cite{Surguladze:1990sp} and
the
 fact that
 $\Gamma_{\bar{q}q{\cal O}_h}^{0,h}(p,-p)$ denotes an amputated Green function. 
Exploiting the fact that the coefficient functions
do not depend on the momentum transfer, one can set also $p=0$. In this case, on
the l.h.s. of  Eq.~(\ref{eq::gf}) only the hard parts of the Green
function survive, as the massless tadpoles are set to zero in DRED and
DREG. As before, the superscript $h$ stands for hard contributions.

 The
validity of the approximation $m_h^2=p_h^2\approx 0$ was extensively
studied in the context of the SM and reconfirmed for the case of gluon
fusion at two-loop order in SQCD in Ref.~\cite{Degrassi:2008zj}. 
We expect that this approximation holds also in the case of fermionic Higgs
decays, due to the heavy supersymmetric mass spectrum.

There are two possibilities currently used in the literature for the 
derivation of $\Gamma_{\bar{q}q{\cal O}_h}^{0,h}(0,0)$. The first one is
the direct computation of  the scalar and  vector components of the vertex
function making use of the appropriate projectors. The second one uses
the LET, which relates  the vertex corrections to the $hq\bar{q}$ coupling
 to the quark self-energy
corrections via appropriate derivatives~\cite{Ellis:1975ap,Degrassi:2008zj}.

\subsection{ Direct calculation of the coefficient function $C_{2q}^0$ at NNLO}
Decomposing the Green
function $\Gamma_{\bar{q}q{\cal O}_h}^{0,h}$ into its scalar and
vector components $\Gamma_{\bar{q}q{\cal O}_h; s}^{0,h}$,
$\Gamma_{\bar{q}q{\cal O}_h; v}^{0,h}$ 
 one derives from
Eq.~(\ref{eq::gf}) two linearly independent relations for the
coefficients $C_{2q}^{0}$ and $C_{3q}^{0}$
\begin{eqnarray}
\Gamma_{\bar{q}q{\cal O}_h; s}^{0,h}(0,0)&=&\zeta_2^{(0)}
\zeta_m^{(0)}(C_{2q}^{0}-C_{3q}^{0})\,,
\nonumber\\
\Gamma_{\bar{q}q{\cal O}_h; v}^{0,h}(0,0)&=&\zeta_2^{(0)}C_{3q}^{0}\,.
\end{eqnarray}
In SQCD at the two-loop order there is also an axial contribution to
 $\Gamma_{\bar{q}q{\cal O}_h}^{0,h}(0,0)$, which arises from diagrams where
 a top quark-squark pair is exchanged from a gluino propagator. However, it
 generates  only contributions  ${\cal O}(\alpha_s^4)$ to
 $\Gamma(h\to q\bar{q})$ which are beyond the precision we are interested
 in this paper. 

Finally, using Eqs.~(\ref{eq::letse}) the expression for $C_{2q}^{0}$ 
through two loops reads
\begin{eqnarray}
C_{2q}^{0}&=&\frac{\Gamma_{\bar{q}q{\cal
      O}_h; s}^{0,h}(0,0)}{1-\Sigma_s^{0,h}(0)}
+\frac{\Gamma_{\bar{q}q{\cal O}_h;
    v}^{0,h}(0,0)}{1+\Sigma_v^{0,h}(0)}\,.
\label{eq::c2mssm}
\end{eqnarray}
One can either work in the
$(\phi_1,\phi_2)$ basis, {\it i.e.} generating two sets of vertex
corrections for each Higgs field, or one derives the Feynman rules for
the couplings of quarks and squarks to the light Higgs using the
relation Eq.~(\ref{eq::hmix}).

For the computation of the vertex corrections up to two loops we have
implemented two independent setups. In one of them, the Feynman diagrams
are generate with the help of the program {\tt
  FeynArts}~\cite{Hahn:2000kx}, then 
its output is handled in a self-written {\tt Mathematica} code which
includes the two-loop tensor reduction and the mapping of the vertex
topologies  to the two-loop tadpole ones~\cite{Davydychev:1992mt}. The
last step is 
possible due to the fact, that we neglect the mass of the light Higgs
boson and of the external light quarks. In the second setup, the  Feynman
diagrams are generated with the program {\tt QGRAF}~\cite{Nogueira:1991ex}, 
and further processed with  {\tt q2e} and {\tt
  exp}~\cite{Harlander:1997zb,Seidensticker:1999bb}. The
reduction  of various vacuum integrals to the master integral was
 performed by a self-written {\tt  FORM}~\cite{Vermaseren:2000nd} routine.

\subsection{ \label{sec::letsusy} LET derivation of the coefficient
  function $C_{2q}^0$ at NNLO} 
The connection between the coefficient functions $C_1^0, C_{2q}^0$
and the decoupling coefficients $\zeta_s^0, \zeta_m^0 $ or
equivalently  $\Pi^{0,h}(0), \Sigma_s^{0,h}(0)$ and  $\Sigma_v^{0,h}(0)$
was extensively studied in the context of the SM. The validity of the
LET was verified up to three-loop order in QCD~\cite{Chetyrkin:1997un}.
In the framework of the MSSM, however, the derivation of LET is much more
involved due to the presence of the two Higgs doublets and of many
massive particles and mixing angles. The applicability of LET in SQCD at
two-loop order was verified only very recently. Namely, the relationship
between the 
coefficient function $C_1$ and  the hard part of the transverse gluon
polarization function $\Pi^{0,h}(0)$ has been established in
Ref.~\cite{Degrassi:2008zj}. Furthermore, the leading two-loop
contributions to the effective 
bottom Yukawa couplings have been derived from the scalar part of the
bottom quark self-energy in Ref.~\cite{Noth:2010jy}.
It is one of  the aims of this paper  
to verify the relationship
between the coefficient function $C_{2q}$ and the hard part of the scalar
and vector contributions to the quark self-energy.

For our calculation it is very convenient to work in
$(\phi_1,\phi_2)$ basis, which means that we have to decompose the Green
functions  according to Eq.~(\ref{eq::hmix})
\begin{eqnarray}
\Gamma_{\bar{q}q{\cal O}_h; a}^{0,h}(0,0)=-\sin\alpha\,
\Gamma_{\bar{q}q{\cal O}_{\phi_1}; a}^{0,h}(0,0)
+ \cos\alpha\, \Gamma_{\bar{q}q{\cal O}_{\phi_2}; a}^{0,h}(0,0)\,,\quad a=s,v\,.
\end{eqnarray}
Similarly, the coefficient function can be written as follows
\begin{eqnarray}
C_{2q}^{0}&=& -\sin\alpha \, C_{2q,\phi_1}^{0}+\cos\alpha \,
 C_{2q,\phi_2}^{0}\,.
\label{eq::c2_12}
\end{eqnarray}
 Applying the LET\footnote{The gauge-fixing condition for SQCD is
   independent of the vacuum expectation values $v_{1,2}$, so that LET
   holds in its trivial form~\cite{Pilaftsis:1997fe}.}  to the
 individual components 
 $\Gamma_{\bar{q}q{\cal O}_{\phi_i}; a}$ we get
\begin{eqnarray}
\Gamma_{\bar{q}q{\cal O}_{\phi_i};\, s}^{0,h}(0,0)&=&\frac{1}{m_b}\frac{\partial
  }{\partial\phi_i}\bigg[m_b(1-\Sigma_s^{0,h}(0))\bigg]\bigg|_{\phi_i=v_i}\equiv\frac{1}{m_b} \hat{D}_{q,\phi_i}  \bigg[m_b(1-\Sigma_s^{0,h}(0))\bigg]
\,,
\nonumber\\
\Gamma_{\bar{q}q{\cal O}_{\phi_i};\, v}^{0,h}(0,0)&=&\frac{\partial
  }{\partial\phi_i}\bigg[-\Sigma_v^{0,h}(0)\bigg]\bigg|_{\phi_i=v_i}\equiv \hat{D}_{q,\phi_i}  \bigg[-\Sigma_v^{0,h}(0)\bigg]
\,,
\label{eq::let}
\end{eqnarray}
with $ i=1,2$ and  $a=v,s$.
As we are considering only the hard parts of the above Green
functions no complication related to the occurrence of infrared
divergences is encountered. 
In practice, it is convenient to express the operators
$\hat{D}_{q,\phi_i}$ introduced in Eq.~(\ref{eq::let}) in terms of
derivatives w.r.t.\ masses and mixing angles. This can be achieved 
using the field-dependent definition of the parameters, in our case
quark and squark masses and squark mixing angles~\cite{Brignole:1991pq}.

The formulas  derived up to now are valid for a generic light quark
flavour $q$. However, for 
phenomenological applications  the decay channel $h\to b\bar{b}$ is the
most important one. The explicit expressions for the operators 
$\hat{D}_{b,\phi_i}$  can be easily derived
from the Eqs.~(11) and (12)  in Ref.~\cite{Degrassi:2008zj}. 
 We quote them here for completeness and to fix our normalization:
\begin{eqnarray}
\hat{D}_{b,\phi_1}&=&\frac{1}{\cos\beta}( m_b   A_b {\cal F}_b +m_b {\cal G}_b)
-\frac{1}{\sin\beta} m_t
\mu_{\rm SUSY}\sin 2 \theta_t {\cal F}_t\,,\nonumber\\
\hat{D}_{b,\phi_2}&=& \frac{1}{\cos\beta} (-m_b\mu_{\rm SUSY} {\cal F}_b)
+ \frac{1}{\sin\beta} (m_t A_t \sin 2 \theta_t {\cal F}_t
+2 m_t^2 {\cal G}_t)\,, \quad \mbox{with}\nonumber\\
{\cal F}_b&=& \frac{2}{\Msbu^2-\Msbd^2} (1-\Sqq)
\frac{\partial}{\partial \Sq}\,,\quad {\cal G}_b = \frac{\partial}{\partial m_b}\,,\nonumber\\
{\cal F}_t&=& \frac{\partial}{\partial \Mstu^2}-
\frac{\partial}{\partial \Mstd^2}
+\frac{2}{\Mstu^2-\Mstd^2} \frac{ (1-\Stq)}{\St}
\frac{\partial}{\partial \St}\,,\nonumber\\
{\cal G}_t&=&\frac{\partial}{\partial \Mstu^2}
+\frac{\partial}{\partial \Mstd^2}
+\frac{\partial}{\partial m_t^2}\,.
\label{eq::derivop}
\end{eqnarray}
In the above formulas we keep only the terms that do not vanish
 in the limit $m_b\to 0$.
 
As is well known, in  Eqs.~(\ref{eq::let})  one has  first
to apply the derivative operators $\hat{D}_{q,\phi_i}$ and afterwards
perform the renormalization. For simplicity of the notation we suppress
the superscript $(0)$, labeling bare quantities. We  checked
explicitly at the diagram
level  that Eqs.~(\ref{eq::let}) hold through two loops. 
The computation of the two-loop 
diagrams contributing to $\Sigma_a^{0,h}(0)$ goes along the same 
line as that for 
$\Gamma_{\bar{q}q{\cal O}_{\phi_i};\, a}^{0,h}(0,0)$. The exact 
results together with  
 few expansions for special mass hierarchies can be
 found in Ref.~\cite{Bednyakov:2007vm,Bauer:2008bj}.

Let us mention at this point that for large values of $\tan\beta$ the
dominant contribution  to the coefficient function $C_{2q}$ is contained in the
first term in Eq.~(\ref{eq::c2mssm}). The $\mu_{\rm SUSY}\tan\beta$-enhanced
contributions are implicitly resummed in Eq.~(\ref{eq::c2mssm}), through
the presence of the denominator $1-\Sigma_s^{0,h}(0)$, which contains
 contribution of the form $\alpha_s^n\mu_{\rm SUSY}\tan\beta$.
In the framework of
LET  the $\mu_{\rm SUSY}\tan\beta$-enhanced contributions
 to $\Gamma_{\bar{q}q{\cal O}_h;s}$
 are generated
through the  term proportional to the derivative ${\cal F}_b$ in
 $\hat D_{b,\phi_2}$. Taking into account the
 parametric dependence of $\Sigma_s^{0,h}$ on masses and mixing angles, 
one can easily derive    these  contributions 
 from the terms proportional to $\sin 2\theta_b$ in $\Sigma_s^{0,h}$.\footnote{A similar observation was made in Ref.~\cite{Noth:2010jy}, too.} 
Such contributions have also been  derived in Ref.~\cite{Noth:2010jy} using
the effective Lagrangian approach.
Indeed,  after discarding the additional pieces
comprised in our computation of the coefficient $C_{2q}$, namely the vector part and the rest of the derivative operators in Eq.~(\ref{eq::derivop}),
  we get  good
numerical agreement with the results of Ref.~\cite{Noth:2010jy}.

\subsection{\label{sec::ren} Regularization and renormalization scheme}  
It is well known that the appropriate  regularization scheme for
 the computation of 
 radiative  corrections in  supersymmetric theories is DRED. However,
 the most convenient regularization scheme for the handling  of the 
 dimension four operators at
 higher orders is DREG as discussed in Section~\ref{sec::framework}.
 For the renormalization  we employed two different approaches.
  In one of them we computed the radiative corrections to the 
coefficient functions directly in  DREG. This implies that some
 of the supersymmetric  relations
 between  couplings of quarks and squarks do not hold anymore. More precisely,
 one has to distinguish between the   gluino-quark-squark coupling $\hat{g}_s$ 
 and the gauge coupling
 $g_s$, and   between the Yukawa couplings of Higgs bosons to  
quarks $g_q^{\phi_i}$
 and squarks $g_{\tilde{q}; kr}^{\phi_i}$.  The 
relationships between the different couplings are necessary only at
 the one-loop order
and they are well known since long time~\cite{Martin:1993yx}.\\
In the second approach,  we performed the two-loop computation of the
coefficient functions
 in the DRED scheme and afterwards converted the results into the DREG
 scheme using  
   the two-loop translation relations  for the quark masses
 and strong couplings 
defined in the full~\cite{Mihaila:2009bn} and effective
 theory~\cite{Harlander:2006rj},
 and Eqs.~(\ref{eq::let}).     
As a consistency check, we explicitly verified that the results 
obtained with the two methods agree. 

For the renormalization of the divergent parameters we used the on-shell
 scheme for the gluino and bottom
 squark masses and mixing angle and the minimal subtraction scheme \msbar{}
 or \drbar{}
 for the strong coupling and
 the bottom quark mass.  The renormalization of the trilinear coupling
 $A_b$ was performed
 implicitly through the use of relation
 Eq.~(\ref{eq::mixangle}).
The explicit formulas for the one-loop counterterms are 
well-known in the literature (see for example Ref.~\cite{Pierce:1996zz}). 
   
The complete two-loop  results for the SQCD corrections to the
coefficient function 
$C_{2q}$ discussed in this section are too lengthy to be given
here. They are available in {\tt   MATHEMATICA} 
format upon request from the authors.
For further applications, we  also
 provide results for the case where all parameters are renormalized minimally.\\

In principle, the results obtained in this section can be
 easily generalized to the
 heavy Higgs decays taking into account the necessary changes in
the Yukawa couplings as given in Table~\ref{tab::yukawa_coeff}. However,
 the application of
 effective theory
formalism introduced in  Section~\ref{sec::nnlo} is not justified in
this case, {\it i.e.~} the condition $M_H\ll M_t\,,M_{\rm SUSY}$
 does not hold anymore.

\section {Numerical results}
In this section we study the phenomenological implications of the
two-loop corrections to the coefficient function $C_{2q}$ on the
 Higgs decay width $\Gamma(h\to b\bar{b})$. The  SM input parameters are the
strong coupling constant at the Z-boson mass scale
$\alpha_s(M_Z)=0.1184$~\cite{Bethke:2009jm}, the top quark pole mass
$M_t=173.1$~GeV~\cite{:2009ec} and the running bottom quark mass in the
\msbar{} scheme $m_b(m_b)=4.163$~GeV~\cite{Chetyrkin:2009fv}. For the 
supersymmetric mass spectrum we adopted the corresponding values of
the ``small $\alpha_{\rm eff}$''
and ``gluophobic'' scenarios as defined 
in Ref.~\cite{Carena:2002qg}.\footnote{We used the tree-level formulas to
  derive the mass eigenvalues for squark fields.}\\
For the running of  the strong coupling constant within QCD we use the {\tt
  Mathematica}  package {\tt RunDec}~\cite{Chetyrkin:2000yt}. For the
evaluation of the  
strong coupling constant within the six-flavour SQCD and the \drbar{}
scheme we follow Ref.~\cite{Bauer:2008bj}.\footnote{For a more detailed
  discussion see Ref.~\cite{Bauer:2008bj} and the references
  therein.}\\
An important ingredient for the computation of the decay
 width $\Gamma(h\to b \bar{b})$ is the
effective mixing angle of the neutral Higgs sector $\alpha_{\rm
  eff}$~\cite{Heinemeyer:2000fa} that takes into account the radiative
corrections to the Higgs propagator. In practical applications one
replaces the tree-level mixing angle $\alpha$ defined in
Eq.~(\ref{eq::h_alpha}) with $\alpha_{\rm eff}$.
This can be computed in perturbation theory 
from the knowledge of the radiative corrections to
the self-energy matrix of the neutral Higgs doublet
$\hat{\Sigma}_{\phi_1},\,\hat{\Sigma}_{\phi_2},\,\hat{\Sigma}_{\phi_1\phi_2} $,
\begin{eqnarray}
\tan\alpha_{\rm
  eff}=\frac{-(M_A^2+M_Z^2)\sin\beta\cos\beta-
\hat{\Sigma}_{\phi_1\phi_2}}{M_Z^2\cos^2\beta+M_A^2
\sin^2\beta-M_h^2-\hat{\Sigma}_{\phi_1}}\,, 
\label{eq::h_alpha_eff}
\end{eqnarray}
where $M_h$ stands for the on-shell mass of the light Higgs boson.
 For the numerical analyses we implemented 
the exact two-loop results from Ref.~\cite{Degrassi:2001yf}.\footnote{Very recently the three-loop SQCD corrections to $M_h$ have been computed~\cite{Kant:2010tf}. However, they are valid only for specific mass hierarchies.}

\begin{figure}
  \begin{center}
    \begin{tabular}{c}
      \epsfig{figure=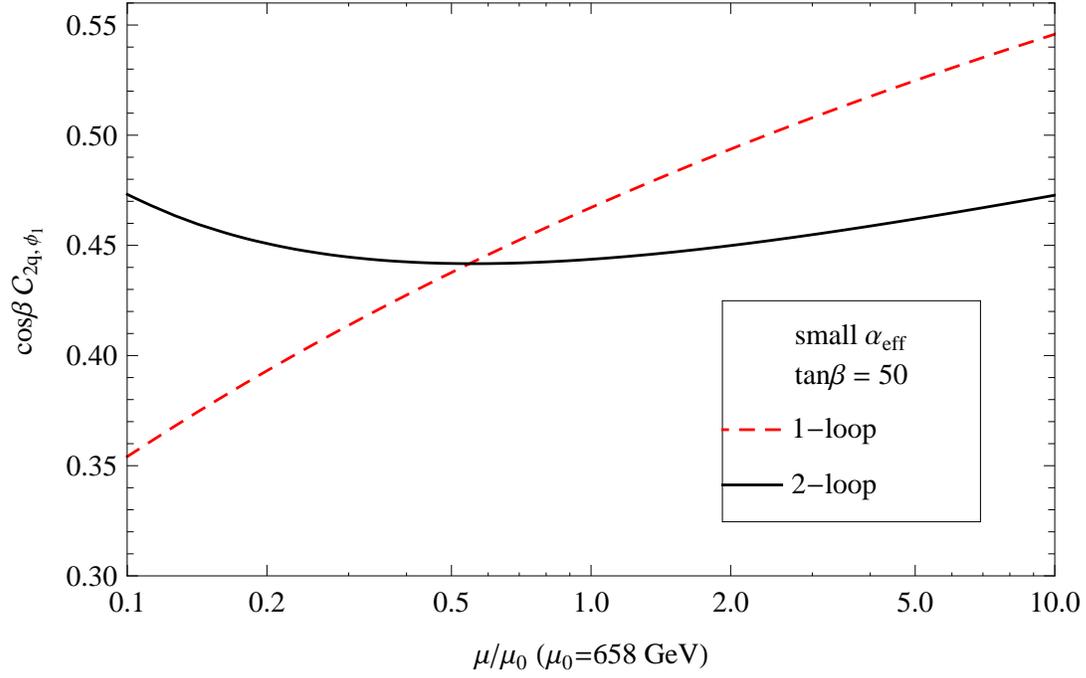,width=.97\textwidth}
      \\(a)\\
      \epsfig{figure=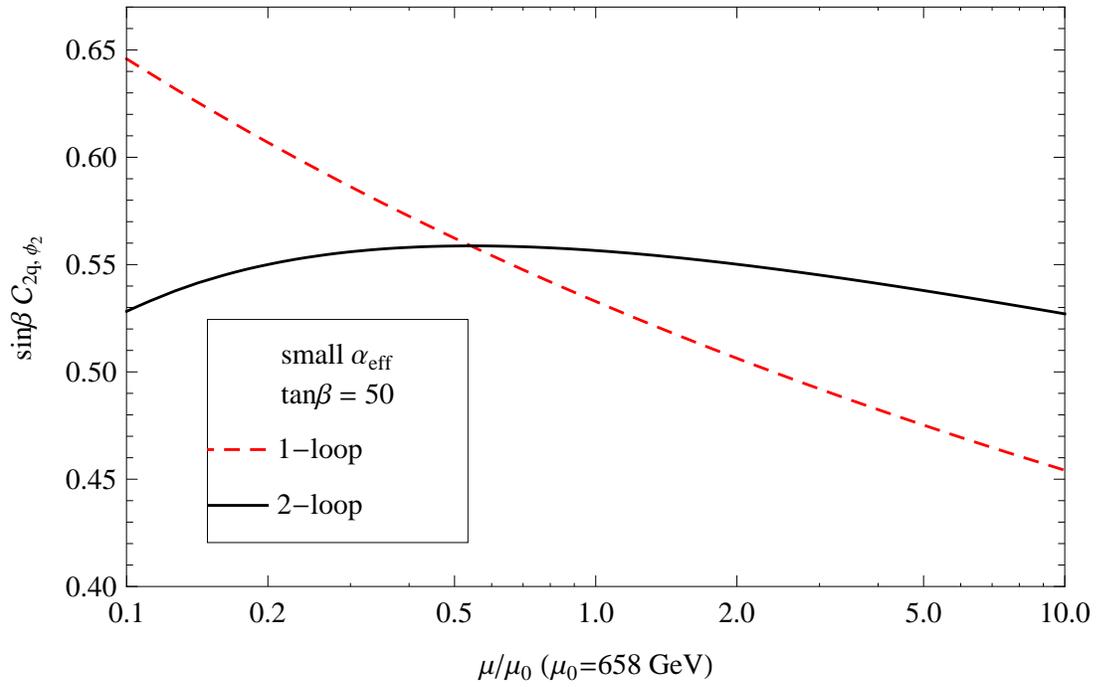,width=.97\textwidth}
      \\(b)
    \end{tabular}
    \parbox{14.cm}{
      \caption[]{\label{fig::muren_alp}\sloppy
        The
        renormalization scale  dependence of the coefficient
        functions ${\cal C}_{2q,\phi_1}\cos\beta$ (a) and
        ${\cal C}_{2q,\phi_2}\sin\beta$ (b)  is depicted at one- and two-loop
        order in the ``small $\alpha_{\rm eff}$'' scenario.
        }}
  \end{center}
\end{figure}

In Fig.~\ref{fig::muren_alp} we show separately the renormalization scale
dependence of the coefficient functions ${\cal C}_{2q,\phi_1}$ and
${\cal C}_{2q,\phi_2}$, which describe the effective Yukawa couplings of the
neutral Higgs fields $\phi_1$ and $\phi_2$. We choose   the ``small $\alpha_{\rm
  eff}$'' scenario with $\tan\beta=50$, $M_A=300$~GeV and evaluate 
$\alpha_{\rm eff}$ 
 with the tree-level formula in order to avoid additional scale
 dependence. The dashed 
and solid lines correspond to the one- and two-loop results, respectively. 
 As can be
read from the figure, at the one-loop order  the scale dependence amounts
to about $50$\% when the renormalization scale is varied around the
average value $\mu_0=(\Msbu+\Msbd+\Mgl)/3\simeq 658$~GeV by a factor 
$10$. At the two-loop order the variation with the renormalization
scale is significantly improved. The remaining scale dependence is 
below $6$\%. An interesting aspect is the opposite evolution of the two
coefficient functions  with the renormalization scale. This feature is
reflected in a milder scale dependence of the full coefficient function
${\cal C}_{2q}$ of about $4$\% and $0.5$\% at one- and two-loop order,
respectively. However, for low A-boson masses $M_A\le 120$~GeV this numbers
 change to
$27$\% and $5$\%, respectively. As usual, we can interpret the last number as an
estimation of the theoretical uncertainties due to unknown higher 
order corrections. So, the two-loop SQCD corrections are essential for the
 accurate prediction of the decay width $\Gamma(h\to b \bar{b})$.

\begin{figure}
  \begin{center}
    \begin{tabular}{c}
      \epsfig{figure=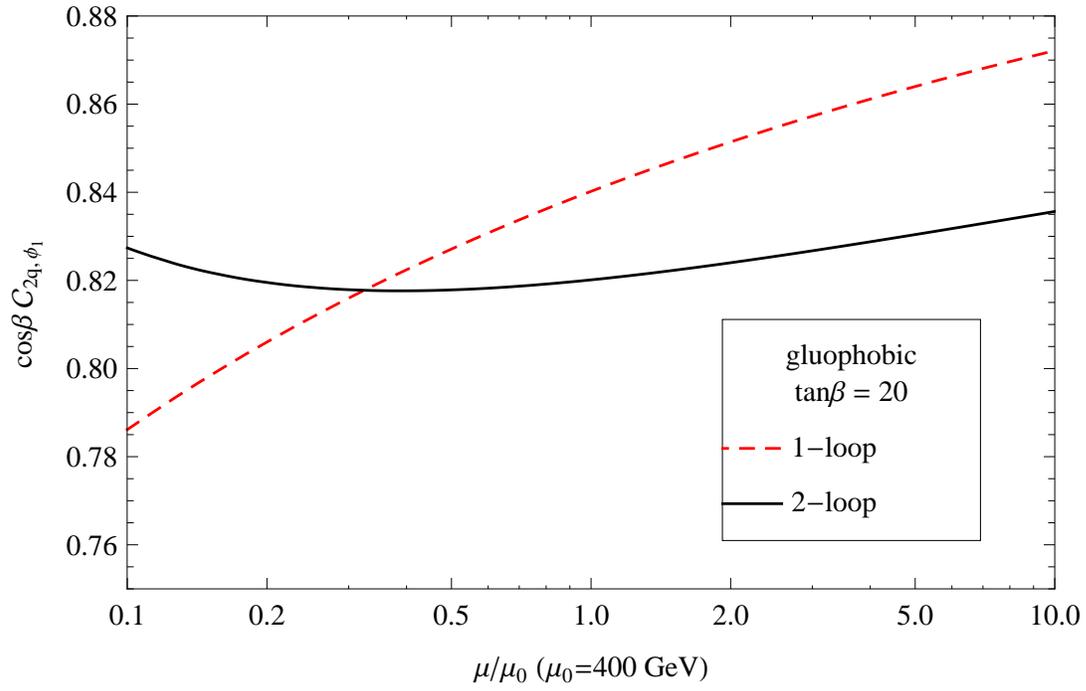,width=.97\textwidth}
      \\(a)\\
      \epsfig{figure=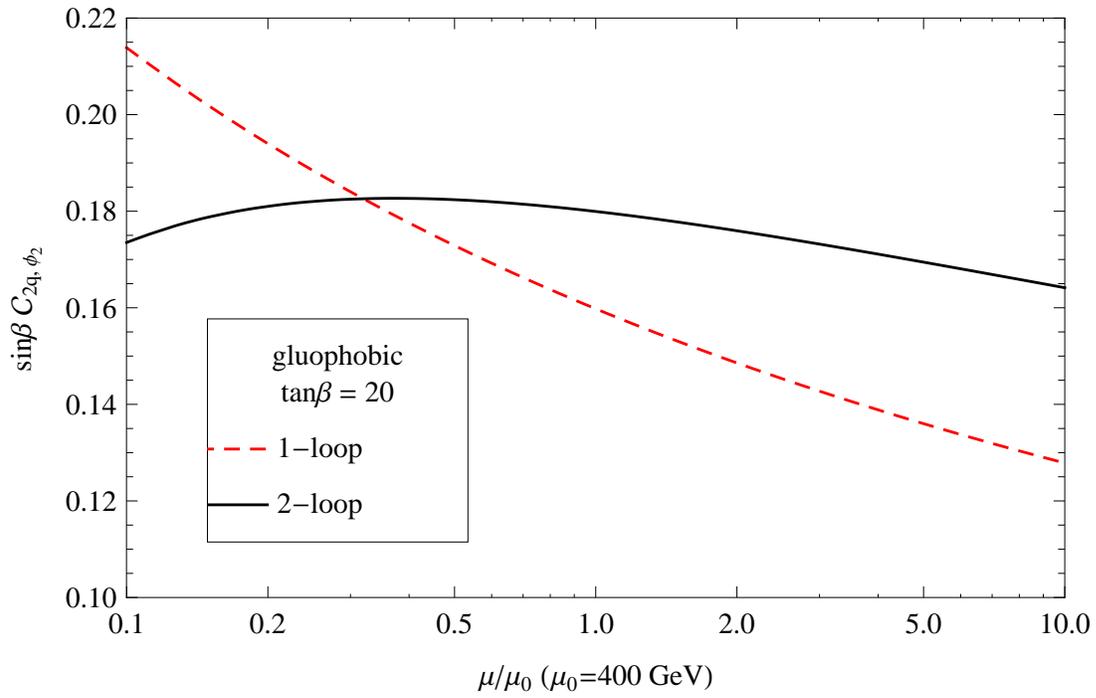,width=.97\textwidth}
      \\(b)
    \end{tabular}
    \parbox{14.cm}{
      \caption[]{\label{fig::muren_gl}\sloppy
        The
        renormalization scale  dependence of the coefficient
        functions ${\cal C}_{2q,\phi_1}\cos\beta$ (a) and
        ${\cal C}_{2q,\phi_2}\sin\beta$ (b)  is depicted at one- and two-loop
        order in the ``gluophobic'' scenario.
        }}
  \end{center}
\end{figure}

In Fig.~\ref{fig::muren_gl} we display  the  renormalization scale
dependence for the
``gluophobic'' scenario, where we fixed the value of $\tan\beta$  to
$20$, $M_A=300$~GeV and maintain the same convention for the lines.
 A similar behaviour as for the ``small $\alpha_{\rm eff}$'' scenario is
 observed. However, in this case the 
coefficient function ${\cal C}_{2q,\phi_2}$  has a much stronger scale
dependence at the one-loop order than the coefficient ${\cal C}_{2q,\phi_1}$.
 The scale variation of the full coefficient function ${\cal C}_{2q}$ sums up to
$1.5$\% and $0.2$\% for $M_A=300$~GeV at the one- and two-loop order,
respectively. However for 
 $M_A\le 120$~GeV the scale variation amounts to $8$\% and $1.5$\% at one- and
 two-loop order, respectively, which shows the importance of the two-loop SQCD 
corrections 
for this region of the parameter space.\\
We can conclude that for phenomenological analyses 
the choice of the renormalization scale around $\mu_0$ ensures
 small radiative corrections and a good convergence of the perturbative scale.
 In the following, we set $\mu\simeq \mu_0$ for the computation of 
 the SQCD corrections to the coefficient functions ${\cal C}_{2q}$ and the decay
 width $\Gamma(h\to b\bar{b})$.

\begin{figure}
  \begin{center}
    \begin{tabular}{c}
      \epsfig{figure=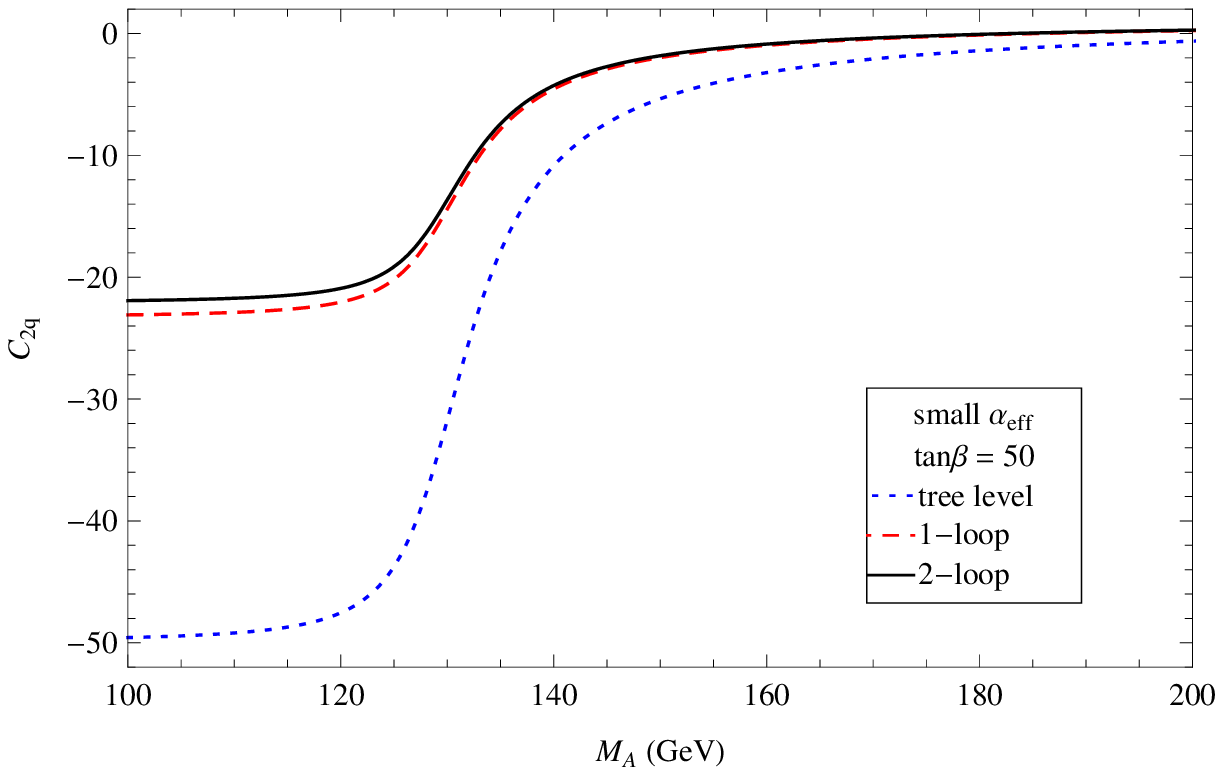,width=.93\textwidth}
      \\(a)\\[1em]\quad
      \epsfig{figure=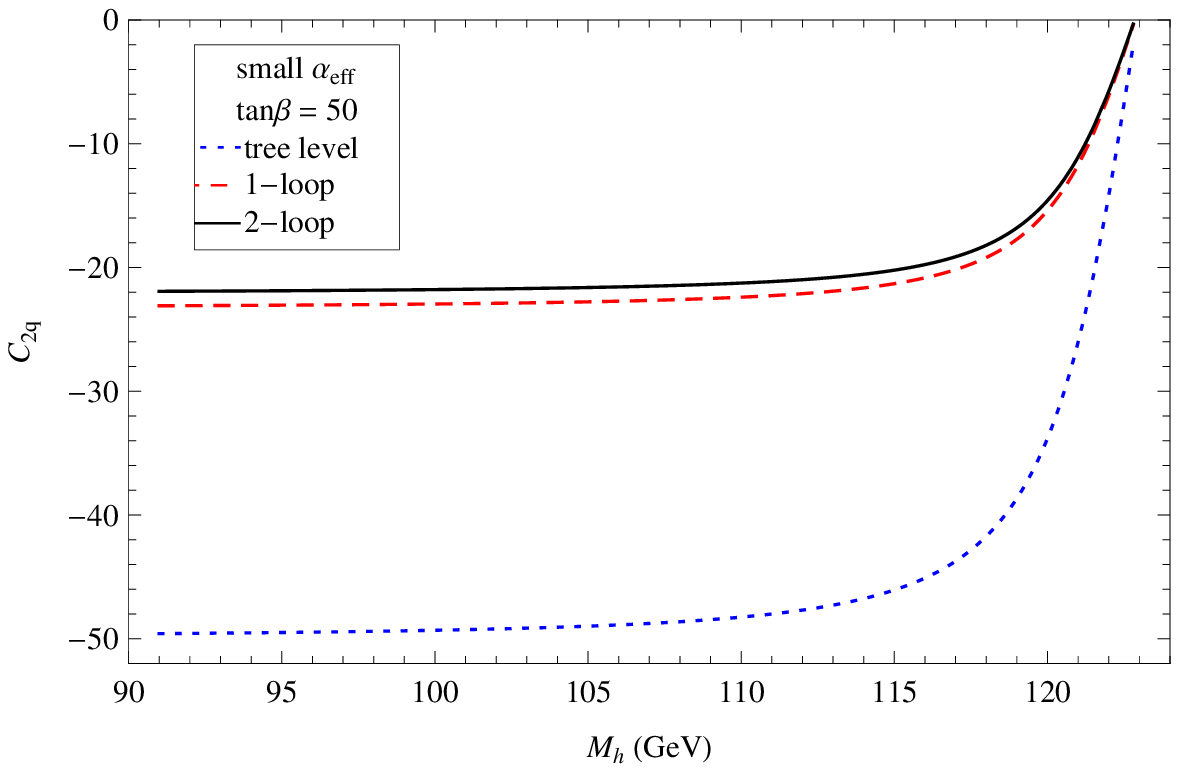,width=.95\textwidth}
      \\(b)
    \end{tabular}
    \parbox{14.cm}{
      \caption[]{\label{fig::ma_mh}\sloppy
        The coefficient function $C_{2q}$ as a function of $M_A$ (a) and
        $M_h$ (b)  is shown at one- and two-loop
        order in the ``small $\alpha_{\rm eff}$'' scenario.
        }}
  \end{center}
\end{figure}

In Fig.~\ref{fig::ma_mh} we depict the dependence of the full coefficient
 function 
$C_{2q}$ on the A boson mass $M_A$ (a) and the light Higgs boson mass
$M_h$ (b) for the ``small $\alpha_{\rm eff}$'' scenario. We set $\tan\beta=50$
 and
 vary the mass of the A boson between $100$~GeV\,$\le M_A\le 200$~GeV. For the
 evaluation of the Higgs boson
 mass and the effective mixing angle $\alpha_{\rm eff}$ we employed the
 two-loop SQCD results~\cite{Degrassi:2001yf}. As can be seen from 
 Fig.~\ref{fig::ma_mh}(a) for low $M_A$ values there are large one-loop
 corrections of about $60$\% of the tree-level values. They originate from
the  large corrections to the scalar part of the quark self-energy, that 
are actually resummed through the use of the formula given in the
 Eq.~(\ref{eq::c2mssm}).  This feature is also reflected by the 
relatively small two-loop corrections of about 
$2$\% of the tree-level values. For $M_A$ values larger than $130$~GeV
 one observes a steep
increase of the coefficient function $C_{2q}$ (decrease of the absolute value)
 due to  cancellation of Eq.~(\ref{eq::h_alpha_eff}), that implies
 $\alpha_{\rm eff}\to 0$. In this case,  $C_{2q}$ reaches its minimal
 absolute value. 
From panel (b) one notices a similar steep increase of the coefficient 
function  $C_{2q}$
when $M_h$ reaches its maximal value of about $125$~GeV.
 A similar behaviour is  also observed  for the ``gluophobic'' scenario. So, for
 both scenarios we expect a strong suppression 
of the decay width $\Gamma(h\to b\bar{b})$ 
for large A-boson masses or equivalently
 for $M_h$ close to its maximal value for which $\alpha_{\rm eff} \to 0$.

\begin{figure}
  \begin{center}
    \begin{tabular}{c}
      \epsfig{figure=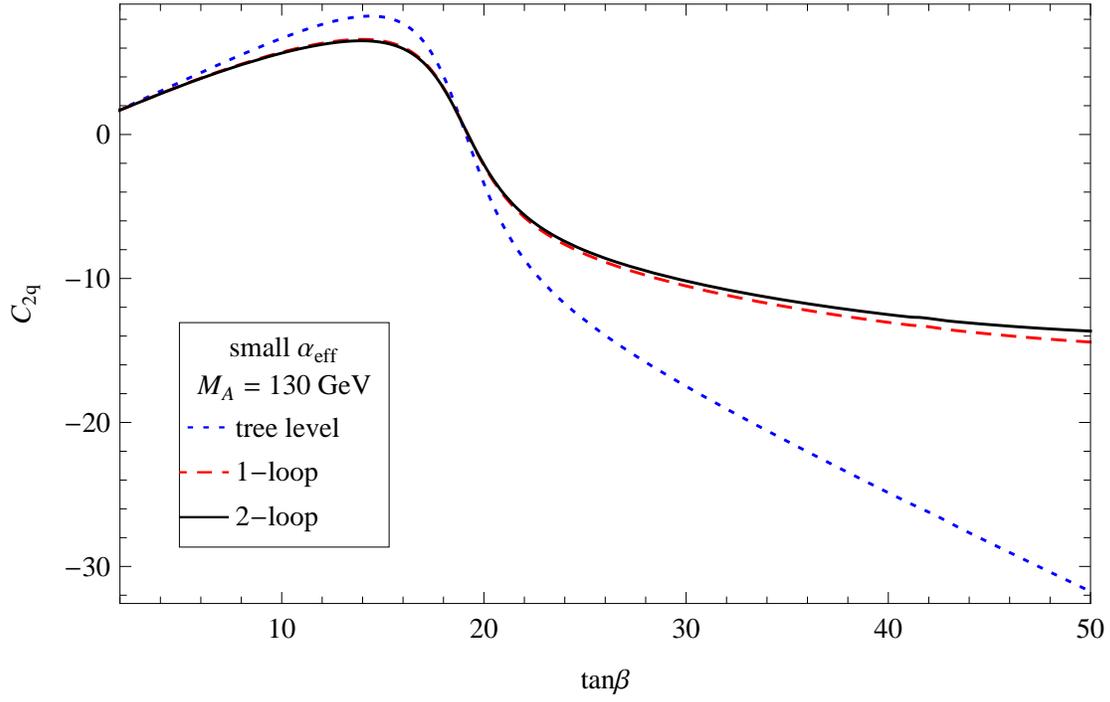,width=.94\textwidth}
      \\(a)\\[1em]
      \epsfig{figure=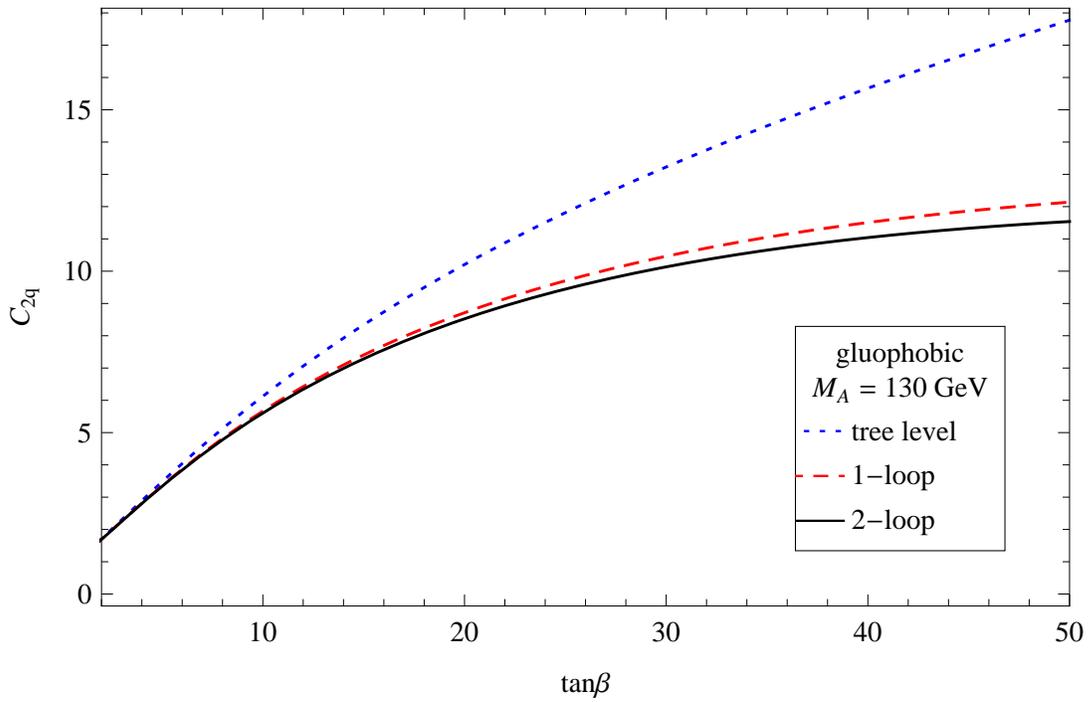,width=.95\textwidth}
      \\(b)
    \end{tabular}
    \parbox{14.cm}{
      \caption[]{\label{fig::tbeta}\sloppy
        The coefficient function $C_{2q}$ as a function of
        $\tan{\beta}$ for $M_A=130$~GeV in (a) the ``small 
$\alpha_{\rm eff}$'' 
         and (b) the ``gluophobic'' scenario.
        }}
  \end{center}
\end{figure}

In Fig.~\ref{fig::tbeta} the dependence of the coefficient function $C_{2q}$ 
on $\tan\beta$ is shown for the ``small $\alpha_{\rm eff}$'' (a) and
 ``gluophobic'' (b)
 scenarios, where the A-boson mass was fixed to $M_A=130$~GeV. As expected,
 we observe 
a significant increase of the magnitude of the radiative corrections
 with the increase of the  
$\tan\beta$ value. At the one-loop order  the radiative corrections amount
 to about  70\% (a) and 
 to 33\% (b) from the tree level values. At the two-loop order they sum up
 to 3\% and 5\%, respectively.

\begin{figure}
  \begin{center}
    \begin{tabular}{c}
      \epsfig{figure=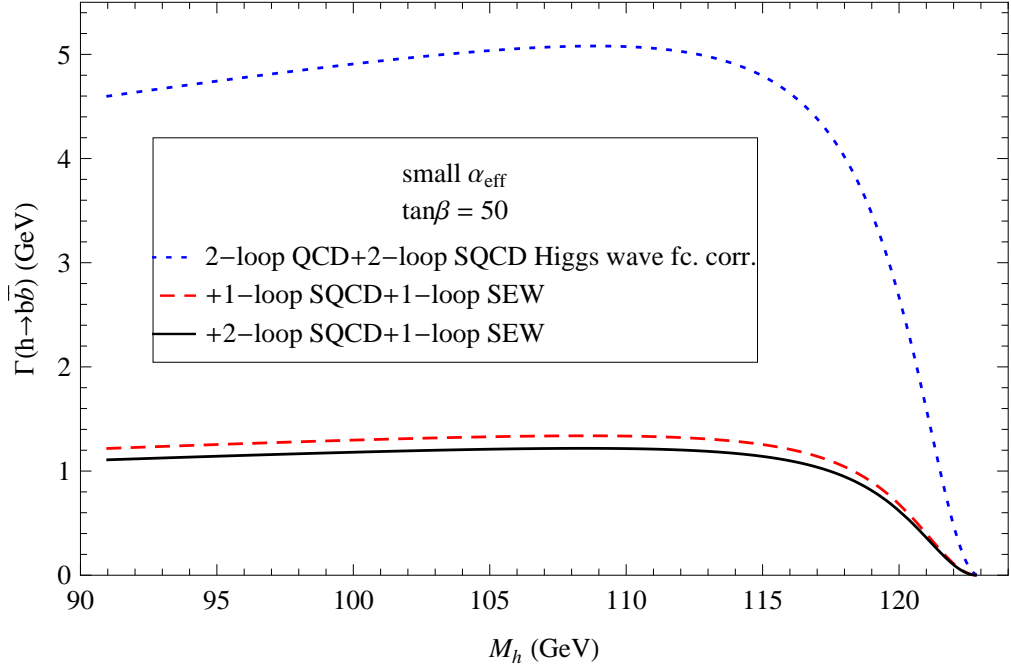,width=.91\textwidth}
      \\(a)\\
      \epsfig{figure=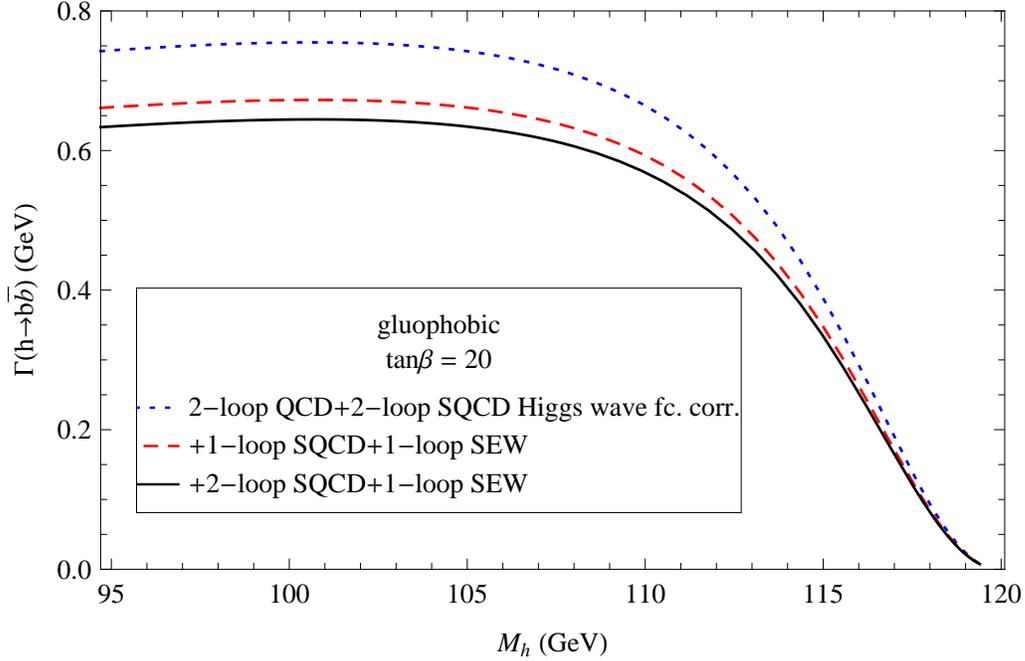,width=.91\textwidth}
      \\(b)
    \end{tabular}
    \parbox{14.cm}{
      \caption[]{\label{fig::gamma}\sloppy  $\Gamma(h\to b\bar{b})$
 for (a) the
``small $\alpha_{\rm eff}$''  and (b) ``gluophobic'' scenario as a function of
 $M_h$. The dotted lines display
 the two-loop QCD and EW corrections together with two-loop corrections to the
 Higgs boson propagator. The dashed and solid lines depict in addition 
the one- and two-loop SUSY-QCD vertex corrections, respectively.          
        }}
  \end{center}
\end{figure}

In Fig.~\ref{fig::gamma} we display the decay width for $h\to b\bar{b}$ as
 a function of the Higgs boson mass $M_h$, 
considering the   ``small $\alpha_{\rm eff}$''(a)  and  ``gluophobic''(b)
 scenarios. We chose $\tan\beta=50$ and $\tan\beta=20$ 
for the case (a) and (b), respectively. The two-loop genuine QCD and EW
 corrections to the process  $h\to b\bar{b}$, as well as the two-loop SQCD
 corrections to the Higgs boson propagator are depicted by the dotted lines.
 More precisely, they are derived from  Eq.~(\ref{eq::gamma_gen}), where
 the coefficient functions
 ${\cal C}_1 \, \mbox{and}\, {\cal C}_2$ are set to their tree-level values. 
The additional SQCD vertex
 corrections parametrized through the coefficient functions 
 ${\cal C}_1 \, \mbox{and}\, {\cal C}_2$ 
 are represented at the one- and two-loop order by the dashed and solid lines,
 respectively. We also take into account the one-loop SEW corrections to
 the coefficient
 function ${\cal C}_2$ and fix their renormalization scale at
 $\mu_{\rm SEW}=(\Mstu+\Mstd+\mu_{\rm SUSY})/15$, for which the two-loop 
 SEW corrections
 become negligible~\cite{Noth:2010jy}.\\
For a relatively light Higgs boson mass $M_h$, the large one-loop
 radiative corrections of about 70\% (a) and 50\% (b)
are still amplified by   mild two-loop corrections that can reach as much
 as about $8$\%  
from  the decay width including QCD corrections even for the selected choice of
 the renormalization scale of SQCD corrections. The large SQCD radiative
 corrections to $\Gamma(h\to b\bar{b})$ have only a relatively small impact 
on the branching ratio $BR(h\to b\bar{b})$ but they can have a large impact
on $BR(h\to \tau^+\tau-)$. For sufficiently large $\tan\beta$ 
and $\mu_{\rm SUSY}$, the measurement of  $BR(h\to \tau^+\tau-)$ 
can provide information about the  distinction between
 the SM and MSSM predictions.\\ 
For a large Higgs boson mass for which $\alpha_{\rm eff}\to 0$, 
the partial decay
 widths for $h\to b\bar{b}$ and $h\to \tau^+\tau^-$ are
   significantly suppressed. In this case the radiative corrections 
(in particular the corrections to the Higgs boson propagator in 
Eq.~(\ref{eq::h_alpha_eff}))  are
 essential for an accurate prediction of  $\Gamma(h\to b\bar{b})$
 and $\Gamma(h\to \tau^+\tau^-)$. Furthermore, the
 $BR(h\to \gamma\gamma)$ will be strongly enhanced, improving the LHC
 prospects of finding a light Higgs. 

\section{\label{sec::concl}Conclusions}
The knowledge of the Higgs boson couplings is essential for its
 searches at the present hadron colliders. 
In this paper we calculate the two-loop corrections 
of   ${\cal O}(\alpha_s^2)$ to the  Yukawa couplings within the MSSM. 
We employed the effective Lagrangian approach under the assumption of large
 top quark and supersymmetric particle masses. We calculate analytically the
 two-loop corrections to  the coefficient function $C_{2q}$ , taking into 
account the complete mass dependence. For large values of $\tan\beta$ the
 radiative corrections need to be resummed, which in  our approach is
performed through the use of the formula given
 in Eq.~(\ref{eq::c2mssm}). Furthermore, we verified at the diagram level
 the applicability of the low-energy theorem for Higgs interactions
 in the framework of the MSSM as stated in Ref.~\cite{Degrassi:2008zj}.

From the phenomenological point of view, the two-loop corrections presented here
reduce significantly the theoretical uncertainties, estimated through the 
variation with the renormalization scale, at the percent level. The two-loop SQCD corrections 
 become sizable for $\tan\beta\ge 20$ and $M_A\le 130$~GeV.


\bigskip
\noindent
{\large\bf Acknowledgements}\\ 
We are grateful to Matthias Steinhauser and 
Konstantin Chetyrkin for enlightening conversations and many valuable comments.
We thank Michael Spira for providing us with 
results  necessary
for the numerical comparison with Ref.~\cite{Noth:2010jy}
and Pietro Slavich for providing us with the analytic result for the
two-loop SQCD corrections to the light Higgs boson mass.
This work was supported by the DFG through SFB/TR~9 ``Computational Particle
Physics''.





\begin{thebibliography}{99}







%
%

\bibitem{Nilles:1983ge}
  H.~P.~Nilles,
  Phys.\ Rept.\  {\bf 110} (1984) 1.\\
  H.~E.~Haber and G.~L.~Kane,
  Phys.\ Rept.\  {\bf 117} (1985) 75.

\bibitem{Carena:2002qg}
  M.~S.~Carena, S.~Heinemeyer, C.~E.~M.~Wagner and G.~Weiglein,
  Eur.\ Phys.\ J.\  C {\bf 26} (2003) 601

\bibitem{Chetyrkin:1996sr}
  K.~G.~Chetyrkin,
  Phys.\ Lett.\  B {\bf 390} (1997) 309

\bibitem{Chetyrkin:1997vj}
  K.~G.~Chetyrkin and M.~Steinhauser,
  Phys.\ Lett.\  B {\bf 408} (1997) 320

\bibitem{Chetyrkin:1996ke}
  K.~G.~Chetyrkin, B.~A.~Kniehl and M.~Steinhauser,
  Phys.\ Rev.\ Lett.\  {\bf 78} (1997) 594;
  Nucl.\ Phys.\  B {\bf 490} (1997) 19

\bibitem{Hall:1993gn}
  L.~J.~Hall, R.~Rattazzi and U.~Sarid,
  Phys.\ Rev.\  D {\bf 50} (1994) 7048;\\
  R.~Hempfling,
  Phys.\ Rev.\  D {\bf 49}, 6168 (1994);\\
  M.~S.~Carena, M.~Olechowski, S.~Pokorski and C.~E.~M.~Wagner,
  Nucl.\ Phys.\  B {\bf 426} (1994) 269

\bibitem{Heinemeyer:2000fa}
  S.~Heinemeyer, W.~Hollik and G.~Weiglein,
  Eur.\ Phys.\ J.\  C {\bf 16} (2000) 139

\bibitem{FeynHiggs}
  {\tt http://www.feynhiggs.de/}

\bibitem{Carena:1999py}
  M.~S.~Carena, D.~Garcia, U.~Nierste and C.~E.~M.~Wagner,
  Nucl.\ Phys.\  B {\bf 577} (2000) 88

\bibitem{Guasch:2003cv}
  J.~Guasch, P.~Hafliger and M.~Spira,
  Phys.\ Rev.\  D {\bf 68} (2003) 115001

\bibitem{Noth:2010jy}
  D.~Noth and M.~Spira,
  [arXiv:1001.1935 [hep-ph]].\\
  D.~Noth and M.~Spira,
  Phys.\ Rev.\ Lett.\  {\bf 101} (2008) 181801

\bibitem{Spiridonov:1984}
  V.~P.~Spiridonov, Report No. INR P-0378, Moscow, 1984.

\bibitem{Chetyrkin:1997un}
  K.~G.~Chetyrkin, B.~A.~Kniehl and M.~Steinhauser,
  Nucl.\ Phys.\  B {\bf 510} (1998) 61

\bibitem{Schroder:2005hy}
  Y.~Schroder and M.~Steinhauser,
  JHEP {\bf 0601} (2006) 051
  K.~G.~Chetyrkin, J.~H.~Kuhn and C.~Sturm,
  Nucl.\ Phys.\  B {\bf 744} (2006) 121

\bibitem{Harlander:2005wm}
  R.~Harlander, L.~Mihaila and M.~Steinhauser,
  Phys.\ Rev.\  D {\bf 72} (2005) 095009


\bibitem{Bednyakov:2007vm}
  A.~V.~Bednyakov,
  Int.\ J.\ Mod.\ Phys.\  A {\bf 22} (2007) 5245

\bibitem{Bauer:2008bj}
  A.~Bauer, L.~Mihaila and J.~Salomon,
  JHEP {\bf 0902} (2009) 037

\bibitem{Bednyakov:2009wt}
  A.~V.~Bednyakov,
  Int.\ J.\ Mod.\ Phys.\  A {\bf 25} (2010) 2437

\bibitem{Ellis:1975ap}
  J.~R.~Ellis, M.~K.~Gaillard and D.~V.~Nanopoulos,
  Nucl.\ Phys.\  B {\bf 106} (1976) 292;
\\
  M.~A.~Shifman, A.~I.~Vainshtein and V.~I.~Zakharov,
  Phys.\ Lett.\  B {\bf 78} (1978) 443;
\\
  M.~A.~Shifman, A.~I.~Vainshtein, M.~B.~Voloshin and V.~I.~Zakharov,
  Sov.\ J.\ Nucl.\ Phys.\  {\bf 30} (1979) 711
  [Yad.\ Fiz.\  {\bf 30} (1979) 1368];
\\
  A.~I.~Vainshtein, V.~I.~Zakharov and M.~A.~Shifman,
  Sov.\ Phys.\ Usp.\  {\bf 23} (1980) 429
  [Usp.\ Fiz.\ Nauk {\bf 131} (1980) 537];
\\
  B.~A.~Kniehl and M.~Spira,
  Z.\ Phys.\  C {\bf 69} (1995) 77;
\\
  W.~Kilian,
  Z.\ Phys.\  C {\bf 69} (1995) 89;
\\
  M.~Spira, A.~Djouadi, D.~Graudenz and P.~M.~Zerwas,
  Nucl.\ Phys.\  B {\bf 453} (1995) 17;

\bibitem{Bardeen:1978yd}
  W.~A.~Bardeen, A.~J.~Buras, D.~W.~Duke and T.~Muta,
  Phys.\ Rev.\  D {\bf 18} (1978) 3998

\bibitem{Gorishnii:1990zu}
  S.~G.~Gorishnii, A.~L.~Kataev, S.~A.~Larin and L.~R.~Surguladze,
  Mod.\ Phys.\ Lett.\  A {\bf 5} (1990) 2703;
  Phys.\ Rev.\  D {\bf 43} (1991) 1633.

\bibitem{Kwiatkowski:1994cu}
  A.~Kwiatkowski and M.~Steinhauser,
  Phys.\ Lett.\  B {\bf 338} (1994) 66
  [Erratum-ibid.\  B {\bf 342} (1995) 455]

\bibitem{Spira:1993bb}
  M.~Spira, A.~Djouadi, D.~Graudenz and P.~M.~Zerwas,
  Phys.\ Lett.\  B {\bf 318} (1993) 347.

\bibitem{Surguladze:1990sp}
  L.~R.~Surguladze and F.~V.~Tkachov,
  Nucl.\ Phys.\  B {\bf 331} (1990) 35

\bibitem{Degrassi:2008zj}
  G.~Degrassi and P.~Slavich,
  Nucl.\ Phys.\  B {\bf 805} (2008) 267

\bibitem{Hahn:2000kx}
  T.~Hahn,
  Comput.\ Phys.\ Commun.\  {\bf 140} (2001) 418\\
  T.~Hahn and C.~Schappacher,
  Comput.\ Phys.\ Commun.\  {\bf 143} (2002) 54

\bibitem{Davydychev:1992mt}
  A.~I.~Davydychev and J.~B.~Tausk,
  Nucl.\ Phys.\  B {\bf 397}, 123 (1993)

\bibitem{Nogueira:1991ex}
  P.~Nogueira,
  J.\ Comput.\ Phys.\  {\bf 105} (1993) 279


\bibitem{Harlander:1997zb}
  R.~Harlander, T.~Seidensticker and M.~Steinhauser,
  Phys.\ Lett.\ B {\bf 426} (1998) 125,

\bibitem{Seidensticker:1999bb}
  T.~Seidensticker, hep-ph/9905298

\bibitem{Vermaseren:2000nd}
  J.~A.~M.~Vermaseren,
  arXiv:math-ph/0010025

\bibitem{Pilaftsis:1997fe}
  A.~Pilaftsis,
  Phys.\ Lett.\  B {\bf 422} (1998) 201

\bibitem{Brignole:1991pq}
  A.~Brignole, J.~R.~Ellis, G.~Ridolfi and F.~Zwirner,
  Phys.\ Lett.\  B {\bf 271} (1991) 123.
  J.~R.~Ellis, G.~Ridolfi and F.~Zwirner,
  Phys.\ Lett.\  B {\bf 262} (1991) 477.

\bibitem{Martin:1993yx}
  S.~P.~Martin and M.~T.~Vaughn,
  Phys.\ Lett.\ B {\bf 318} (1993) 331

\bibitem{Mihaila:2009bn}
  L.~Mihaila,
  Phys.\ Lett.\  B {\bf 681} (2009) 52

\bibitem{Harlander:2006rj}
  R.~Harlander, P.~Kant, L.~Mihaila and M.~Steinhauser,
  JHEP {\bf 0609} (2006) 053

\bibitem{Pierce:1996zz}
  D.~M.~Pierce, J.~A.~Bagger, K.~T.~Matchev and R.~j.~Zhang,
  Nucl.\ Phys.\  B {\bf 491} (1997) 3

\bibitem{Bethke:2009jm}
  S.~Bethke,
  Eur.\ Phys.\ J.\  C {\bf 64} (2009) 689

\bibitem{:2009ec}
    [Tevatron Electroweak Working Group and CDF Collaboration and D0 Collab],
  arXiv:0903.2503 [hep-ex].

\bibitem{Chetyrkin:2009fv}
  K.~G.~Chetyrkin, J.~H.~Kuhn, A.~Maier, P.~Maierhofer, P.~Marquard, M.~Steinhauser and C.~Sturm,
  Phys.\ Rev.\  D {\bf 80} (2009) 074010

\bibitem{Chetyrkin:2000yt}
  K.~G.~Chetyrkin, J.~H.~Kuhn and M.~Steinhauser,
  Comput.\ Phys.\ Commun.\  {\bf 133} (2000) 43.

\bibitem{Degrassi:2001yf}
  G.~Degrassi, P.~Slavich and F.~Zwirner,
  Nucl.\ Phys.\  B {\bf 611} (2001) 403


\bibitem{Kant:2010tf}
  P.~Kant, R.~V.~Harlander, L.~Mihaila and M.~Steinhauser,
  arXiv:1005.5709 [hep-ph].

\end{thebibliography}
\end{document}